\documentstyle[aps,pra]{revtex}
\begin{document}
\newcommand {\nc} {\newcommand}
\nc {\ve} [1] {\mbox{\boldmath $#1$}}
\nc {\la} {\mbox{$\langle$}}
\nc {\ra} {\mbox{$\rangle$}}
\begin{center}
{\Large {\bf MAGNETIC FIELD STIMULATED TRANSITIONS OF EXCITED}}
\end{center}
\begin{center}
{\Large {\bf STATES IN FAST MUONIC HELIUM IONS}}
\end{center}
\begin{center}
{\bf Vladimir S. Melezhik$^{a,b}$ and Peter Schmelcher$^c$}
\end{center}
\begin{center}
{ $^a$ Joint Institute for Nuclear Research, Dubna, \\
Moscow Region 141980, Russian Federation\\
$^b$ Physique Nucl\'{e}aire Th\'{e}orique et Physique
Math\'{e}matique, \\
C.P. 229, Universit\'{e} Libre de Bruxelles, \\
B 1050 Brussels, Belgium \\
$^c$ Theoretische Chemie, Institut f\"ur Physikalische Chemie,\\
Universit\"at Heidelberg, Im Neuenheimer Feld 253, \\
69120 Heidelberg, Germany}
\end{center}

\begin{abstract}
It is shown that one can stimulate, by using the present-day laboratory
magnetic fields, transitions between the $lm$ sub-levels of fast
$\mu He^+$ ions formating in muon catalyzed fusion. Strong  fields
also cause the self-ionization from highly excited states of
such muonic ions. Both effects are the consequence of the
interaction of the bound muon with the oscillating field of the Stark
term coupling the center-of-mass and muon motions of the $\mu He^+$ ion due
to the non-separability of the collective and internal variables
in this system. The performed calculations show a possibility to
drive the population of the $lm$ sub-levels by applying a
field of a few $Tesla$, which affects the reactivation rate and is
especially important to the $K\alpha$ $x$-ray production in
muon catalyzed fusion. It is also shown that the $2s-2p$ splitting in
$\mu He^+$ due to the vacuum polarization slightly decreases the stimulated
transition rates.
\\
PACS number(s): 36.10.Dr, 32.60.+i, 32.80.Dz, 31.15.-p
\end{abstract}

\section{Introduction}

We show that using present-day laboratory
magnetic fields may change essentially the population of the $lm$
sub-levels of the excited states ($n\geq 2$) of the fast muonic ions $\mu
He^{+}$ formating in muon catalyzed fusion ($\mu CF$)[1-4].
This may affect the muon stripping and, especially, the intensities
of the $x$-rays from the ions. Stronger  fields also cause the
self-ionization\cite{5,6} of the highly excited states of the $\mu
He^{+}$ ions.

Both effects are consequences of the non-separability of the collective
and internal degrees of freedom of the charged two-body system ($\mu
He^{+}$) in the presence of an external magnetic field\cite{7}. 
The coupling between the center-of-mass (CM) and light particle (muon)
motions is described by an oscillating Stark term which is proportional to the
mass ratio $m_{\mu}/M$ ($m_{\mu}$ and $M$ are the muon and the
$\mu He^{+}$ mass respectively), the CM-momentum (see below) and the
magnetic field strength $B$\cite{7}. A rather strong coupling of the CM and
muon motions due to the essential non-adiabaticity ($m_{\mu}/M \simeq
1/40$) and the high CM-velocity ($v\simeq 6v_0=6\alpha c,
\alpha\simeq 1/137$) makes the system $\mu He^{+}$, formed in $\mu CF$,
sufficiently sensitive to the external magnetic field.
Thus, as it will be shown below, one can drive the sub-level populations
$P_{nlm}$ of the fast $\mu He^{+}$ in excited states $n\geq 2$
by applying magnetic fields of a few $Tesla$.  However, from the first glance
the obtained result looks quite unexpected, because the muonic atomic unit
of the magnetic field, $B_0=(e/\hbar)^3m_{\mu}^2c \simeq 1.01 \cdot
10^{14}G$, is extremely large, exceeding the existing present-day
laboratory fields by nine orders of magnitude.

In section II we describe our time-dependent approach for
treating the evolution of the two-body charged system in an external
magnetic field. The calculations of the transition rates between the
sub-levels $lm$ of the fast $\mu He^{+}$ ion in the presence of a magnetic
field are discussed in section III. Section IV is devoted to
a discussion of the self-ionization process of highly excited states of the fast
$\mu He^{+}$ ion in a strong field. Finally we provide some conclusions
in section V.

\section{Theoretical approach to
the fast muonic helium ion in a magnetic field}

We start our analysis with the transformed Hamiltonian\cite{7,8} of the
moving $\mu He^{+}$ ion in a magnetic field.  The Hamiltonian of the system
$H =H_0+h+ \Delta U$ consists of two terms describing the CM motion
\begin{equation} H_0(\ve{P},\ve{R}) = \frac{1}{2 M} (\ve{P}-
\frac{Q}{2}\ve{B}\times\ve{R})^2
\end{equation}
and internal degrees of freedom
$$
h(\ve{p},\ve{r}) = \frac{1}{2m_{\mu}}(\ve{p} -\frac{e}{2}
\ve{B}\times\ve{r} +\frac{Q}{2}\frac{m_{\mu}^2}{M^2}\ve{B}\times\ve{r})^2
\\
$$
\begin{equation} +
\frac{1}{2M_0}(\ve{p}+[\frac{e}{2}-\frac{Q}{2M}\frac{m_{\mu}}{M}(M+M_0)]
\ve{B}\times\ve{r})^2 -\frac{ze^2}{r}
\end{equation}
as well as the coupling term
\begin{equation}
\Delta U(\ve{P},\ve{R},\ve{r})=\gamma \frac{e}{M}[\ve{B}\times(\ve{P} -
\frac{Q}{2}\ve{B}\times\ve{R})]\ve{r}
\end{equation}
between the CM and internal motions. Here ($\ve{R},\ve{P}$) and
($\ve{r},\ve{p}$) are the canonical coordinate-momentum pairs for the
CM and internal motions, $M_0$ is the helium nuclear mass. The magnetic field
vector is denoted as $\ve{B}$ and oriented along the $z-$axis,
the charges of the nucleus, the ion and the
muon are $-ze$, $Q$ and $e$, respectively and $\gamma = (M_0+zm_{\mu})/M$.

In references\cite{9,10} it was shown that for the one-electron ion $He^+$
the corrections to the CM motion due to the coupling term (3) are negligible in
not too strong fields and at low CM velocities.  Therefore, the CM
can be treated as a pseudoparticle with mass $M$ and unit charge
($Q=-e$)\cite{9,10}, performing the cyclic motion
\begin{equation} \dot{\ve{R}} =
v_{\perp}(\cos \omega t \ve{n}_x - \sin \omega t \ve{n}_y)
\end{equation}
with the frequency $\omega=Q B/M$ in the $xy$-plane orthogonal to the
vector of the magnetic field $\ve{B}$. 
$v_{\perp}$ is the projection of the initial
CM velocity $\ve{v}(t=0)=\dot{\ve{R}}(t=0)$ onto the $xy$-plane.
By using the classical Hamiltonian equations connecting
the CM momentum $\ve{P}$ and the velocity $\dot{\ve{R}}$ of the
center of mass of the $He^+$ ion
in the field $B$\cite{6,7}, one can transform the mixing term (3) to the
form
\begin{equation}
\Delta U(\ve{P},\ve{R},\ve{r}) = \gamma
e[\ve{B}\times(\dot{\ve{R}}+ \frac{e\gamma}{M} \ve{B}\times\ve{r})]\ve{r}
\simeq \gamma e[\ve{B}\times\dot{\ve{R}}]\ve{r} \ ,
\end{equation}
representing an oscillating Stark field (see Eq.(4)).
Thus we get the simplified effective Hamiltonian of the electron motion
$H(\ve{r},\dot{\ve{R}}(t))$, with $\dot{\ve{R}}(t)$ defined by the Eq.(4).
Such an approach has, however,many limitations and in particular fails to
describe correctly a number of interesting phenomena occurring in ion
physics.
In particular for the muonic ion $\mu He^{+}$ in a magnetic field the
coupling (3) between the internal and CM variables is $m_e/m_{\mu}\simeq
200$ times stronger compared to the electronic counterpart $He^{+}$, what
demands an accurate treatment of the non-separability of the system.

For analyzing the time-evolution of the fast $\mu He^{+}$ ions in a
magnetic field we suggest an approach including the non-adiabatic effects
in the problem (all the terms of the order $m_{\mu}/M$ and $m_{\mu}/M_0$ in
Eqs.(1)-(3)).  It is a mixed treatment by a coupled system of equations
describing quantum mechanically the muon degrees of freedom and
treating classically the CM motion.  It is formulated as the
initial-value problem
\begin{equation}
i\frac{\partial}{\partial t}
\psi(\ve{r},t) = H(\ve{P}(t),\ve{R}(t),\ve{r}) \psi(\ve{r},t)
\end{equation}
$$
\psi(\ve{r}, t=0) = \phi_{n'l'm'}(\ve{r})
$$
with the
effective Hamiltonian
$$
H(\ve{P}(t),\ve{R}(t),\ve{r}) =
-\frac{1}{2m}\Delta_{\ve{r}} -\frac{2}{r} +\frac{\gamma'}{2m}
\ve{B}\cdot\ve{L} +\frac{1}{8m}(\gamma '^{2}+\frac{4m}{M}\gamma^2)
[\ve{B}\times\ve{r}]^2
$$
\begin{equation}
- \frac{\gamma}{M}
[\ve{B}\times(\ve{P}(t)-\frac{Q}{2}\ve{B}\times\ve{R}(t))]\ve{r} \ ,
\end{equation}
depending on the parameters $\ve{P}(t)$ and $\ve{R}(t)$ defined by the
classical Hamiltonian equations of motion
\begin{equation}
\frac{d}{d t} \ve{P}_j(t) = -\frac{\partial}{\partial
\ve{R}_j}H_{cl}(\ve{P}(t),\ve{R}(t))\   ,\
\frac{d}{d t} \ve{R}_j(t) =
\frac{\partial}{\partial \ve{P}_j}H_{cl}(\ve{P}(t),\ve{R}(t))\  ,
\end{equation}
(here $\phi_{n'l'm'}(\ve{r})$ are the Coulomb wave functions of the
bound muon in $\mu He^{+}$ without external field,
$\ve{L}$ is the muon angular momentum, $\gamma'
=(M_{0}^2 - z m_{\mu}^2)/M^2$ and $m=m_{\mu}/(1+m_{\mu}/M_0)$) where the
Hamiltonian $H_{cl}$ is determined as
\begin{equation}
H_{cl}(\ve{P},\ve{R}) = H_0(\ve{P},\ve{R}) + \la\psi(\ve{r},t) \mid
h(\ve{p},\ve{r}) +\Delta U(\ve{P},\ve{R},\ve{r})\mid \psi(\ve{r},t)\ra
\end{equation}
by averaging
the initial Hamiltonian (1-3) over the internal variables $\ve{r}$ at every
time moment and simultaneously integration of the coupled Eqs.(6-9).
Here and below we use $-e=\hbar=1$.

The CM coordinate $Z$ is separated and two pairs of the classical
Hamiltonian equations of motion
\begin{eqnarray}
\frac{d}{dt}P_x=\frac{\omega}{2}(P_y-\frac{Q B}{2}X +\gamma B \la x\ra) \  ,
\frac{d}{dt}P_y=\frac{\omega}{2}(-P_x-\frac{Q B}{2}Y +\gamma B \la y\ra) \
,\\ \frac{d}{dt}X=\frac{1}{M}P_x+\frac{\omega}{2}Y -\frac{\gamma\omega}{Q}
\la y\ra \  ,
\frac{d}{dt}Y=\frac{1}{M}P_y-\frac{\omega}{2}X
+\frac{\gamma\omega}{Q} \la x\ra \nonumber
\end{eqnarray}
coupled also with the Schr\"odinger equation(6) via the terms
$\la x \ra=\la\psi(\ve{r},t)\mid x\mid\psi(\ve{r},t)\ra$ and
$\la y \ra=\la\psi(\ve{r},t)\mid y\mid\psi(\ve{r},t)\ra$ need to be
integrated.

Such an approach is analogous to the ones given in
refs.[11-13] suggested for
semiclassically treating the dynamics of molecular processes\cite{12,13}.
It has the property of
conserving the total energy of the system $\mu He^{+}$ and includes
the coupling between the CM and internal variables, which is important for
the problem of a fast muonic ion in a magnetic field due to the essential
non-adiabaticity of the system ($m_{\mu}/M, m_{\mu}/M_0 \sim 1/40$) and
high CM-velocity ($v_{\perp}\simeq 6 \alpha c$).

The time-dependent three-dimensional Schr\"odinger equation (6) is 
integrated by the method developed in Refs.[14-16] simultaneously with the
system of coupled Hamiltonian equations of motion(10).

In principle one can think also of a time-dependent perturbation
theoretical approach in order to describe the sublevel
mixing induced by the oscillating motional electric field.
However, the large center of mass velocity together with
the strong non-adiabaticity in the case of our muonic helium
would certainly require a very careful estimate of the
possible range of validity of such an approach.
For small center of mass velocities of the ``electronic''
$He^{+}$-ion a perturbation theoretical approach for the
classical dynamics of the ion has been developed in
Ref.\cite{28}.

\section{Transitions between sub-levels $lm$ of fast muonic helium ions
  in a magnetic field}

The fast ions $\mu He^{+}$ are formated in $\mu CF$
due to the muon ``sticking'' process to helium
\begin{equation} dt\mu
\rightarrow \mu^{4}He + n \
\end{equation}
partly in excited states $n\geq 2$ with the kinetic energy
$E_{cm}=Mv^2/2\simeq 3.5 MeV$[1-4,14]. At the present time the effect is
rather well investigated both experimentally and theoretically due to its
importance for the efficiency of $\mu CF$ in deuterium-tritium
mixture\cite{3}.  Particularly an essential dependence of the
muon ``stripping'' from the $\mu He^{+}$ ions (characterized by the
reactivation rate $R$\cite{17,18}) on the population $P_{nlm}$ of its
excited states ($nl\neq 1s$)[18-21] was found.  Here we analyze this
parameter in the presence of an external magnetic field.

We have calculated the time evolution of the population
\begin{equation}
P_{nl}(t) = \sum_{m=-l}^{l} \mid
\la\phi_{nlm}(\ve{r})\mid\psi(\ve{r},t)\ra\mid^2
\end{equation}
of the sub-levels $l$ for states $n=2$ and $3$ of $\mu He^{+}$
for the present-day laboratory fields $B$ with strengths of a few $Tesla$. The
maximal value $v_{\perp}=6v_0=6\alpha c$ of the $\mu He^{+}$ ion CM
velocity projection onto the $xy$-plane, perpendicular to the direction of
the magnetic field $\ve{B}$, is chosen to coincide with the initial
velocity of the ion, $v\simeq 6v_0$ ($E_{cm}=Mv^2/2\simeq 3.5 MeV$),
formating in reaction (11).

The coupled Schr\"odinger equation (6) and classical
equations (10) with the initial conditions
\begin{eqnarray}
\psi(\ve{r},t=0)=\phi_{n'l'm'}(\ve{r}) \nonumber  ,\\
P_x(t=0) = M v_{\perp} \ , P_y(t=0) = 0 \  ,\\
X(t=0) = Y(t=0) = 0\ , \nonumber
\end{eqnarray}
were integrated simultaneously with the same step of integration $\Delta t$
over time $t$.
Details of the computational scheme
applied for the integration of the Schr\"odinger equation (6) can be found in
refs.\cite{15,16}.  The grids over $r\in [0,r_m=400a_{0}]$
(250 grid points) and the
angular variables $\{\theta, \phi\}$ (25 grid points) were constructed
according to ref.\cite{15}. The step of
integration over $t$ was chosen as $\Delta t \leq 2\cdot 10^3t_{0}$, what
permitted to keep the accuracy of the evaluated quantities (12) about a few
percents after $10^5-10^6$ steps of integration. Here and below some values
are given in muonic atomic units of
$a_0=\hbar^2/(m_{\mu}e^2)=2.56\cdot 10^{-11} cm$ and
$t_{0}=\hbar^3/(m_{\mu}e^4)=1.17\cdot 10^{-19} s$.

Results of the calculation are presented in Figs.1 and 2.
We have analyzed two cases: the muon is initially in the $2s$ or $3s$
states of $\mu He^{+}$, i.e. the quantum numbers $n'l'm'$ were fixed as
$200$ or $300$ in Eqs.(13).  The obtained data demonstrate that applying
the magnetic fields of the order of a few $Tesla$ stimulates fast transitions
between the $lm$ sub-levels in excited states ($n\geq 2$) of the $\mu
He^{+}$ ions moving with the velocities defined by the energy output in
the fusion reaction (11). These
transitions occur through energy transfer from the CM to the internal muon
motion and become faster with increasing field strength $B$ or the
$v_{\perp}$ component of the initial CM velocity $v$ (see Figs.2) in
agreement with the classical equations
\begin{equation} \frac{d}{dt}E_{cm} =
-\frac{d}{dt}E_{int} = e\gamma (\ve{B}\times\dot{\ve{R}})\dot{\ve{r}}\ ,
\end{equation}
where $E_{cm}$ and $E_{int}$ are the energy of the center of mass
and internal motion\cite{7} respectively.  This equation is the
consequence of the classical Hamiltonian equations of motion of the
two-body charged system in a magnetic field\cite{5,6,7} and shows a
permanent exchange of energy between the CM and muonic degrees of freedom.

The above results were obtained in the nonrelativistic limit for which the
splitting between different $lm$ states is given exclusively by the
interaction of the muonic atom with the magnetic field.  However due to its
compactness the muonic helium has considerable relativistic corrections
which could in principle suppress the above-calculated transition rates.
The measured $2s_{1/2}-2p_{3/2}$ splitting is $\Delta E_{2S-2P} = 1.527
eV$ \cite{21}. The main contribution in the $\Delta E_{2S-2P}$ splitting is
given by the vacuum polarization (VP)
effect $\Delta E_{2S-2P}^{VP} = 1.667 eV$ \cite{22}, which is dominant
among other relativistic and short-range corrections in the muonic helium
\cite{23}.  Muonic ions (atoms) are much more sensitive than electronic ones
to the VP alteration of the Coulomb interaction, because the dimension of
the muonic ion
$a_{0}/z=\hbar^2/(m_{\mu} e^2 z) \simeq 2.6/z\cdot 10^{-11} cm$
is close to the Compton electron wavelength $\lambda_e=\hbar/m_ec\simeq
3.9 \cdot 10^{-11} cm$.

To evaluate the influence of the main relativistic effect on the magnetic
field stimulated $2s-2p$ transitions we have integrated the Eqs.(6)-(10)
with the additional VP potential
\begin{equation}
\Delta U(r) =-\alpha\frac{2z}{3\pi r}\int_{1}^{\infty}\frac{\sqrt{x^2-1}}{x^2}
(1+\frac{1}{2x^2})e^{-2(r/\lambda_e)x} dx \ ,
\end{equation}
including in the Hamiltonian (7) an effective interaction
between the muon and helium nucleus due to the virtual production of a
single $e^+e^-$ pair.  The results presented in Fig.3 demonstrate that the
slowing down of the $2s$ state depopulation due to the relativistic $2s-2p$
splitting becomes considerable for times of the order $t\simeq 10^{-10}
s$ for the chosen parameters of $B$ and $v_{\perp}$.

The calculated rates of the $2s-2p$ and $3s-3p-3d$ mixing exceed,
at least by two order of magnitude, the rates of the resonance muonic
molecule formation ($\sim10^8-10^9 s^{-1}$)[1-4] and are comparable with
other transition rates of the $\mu He^+$ in the process of the deceleration
in a dense deuterium-tritium mixture (transitions due to inelastic
collisions, Auger transitions, Stark mixing\cite{17,18,24,25}). Transitions
between different $n$, stimulated by the driven field of the order of a few
$Tesla$, are much slower then sub-level transitions with the same $n$.
Thus, for low density with suppression of the collisional transitions
such that only the
radiative transitions $nl\rightarrow n'l'$ between some known
states are remained essential\cite{17,18}, one can drive the sub-levels
population $P_{nl}$ by varying the strength of the applied magnetic field.
Different modelling of the $\mu He^{+}$ ion time-evolution in $dt$ mixture
show the strong dependence of the muonic ions $x$-ray yield from the
$2s-2p$ population (especially for $K_{\alpha}$ radiation).  Some
estimations show also an influence of these populations on the reactivation
rate $R$\cite{17,18,26}.  Both parameters, reactivation rate R and
$K_{\alpha}$ lines intensity, are under extensive experimental
investigation so far\cite{3,19,20,27}.  Thus, the experimental achievement
in $\mu CF$ makes it possible to analyze directly the influence of the
driving magnetic field to the $K_{\alpha}$ $x-$ ray production and the
reactivation rates $R$ by measuring these parameters at low densities in
the presence of external magnetic fields.  The possibility to create a
well-defined mixing of the $l$ sub-levels in such kind of experiments looks
especially valuable, because different hypothesis about the $2s-2p$ mixing
have been used so far in modelling the $\mu He^{+}$
time-evolution\cite{17,18} due to a considerable variation
for the estimates of 
the $2s-2p$ Stark mixing by different authors\cite{18},[25-27].

\section{Self-ionization of highly excited fast muonic helium ions 
in strong magnetic fields}

In refs.\cite{5,6} the self-ionization process for ions in the
presence of an external magnetic field has been predicted
and demonstrated by solving the classical equations of motion for Rydberg
states of the one-electron $He^{+}$ ion.
Here we discuss briefly the possibility to ionize the fast highly-excited
$\mu He^{+}$ ion by a strong magnetic field. Actually, the muon kinetic
energy in the direction of the magnetic field $\ve{B}$ can
become large enough
in order to ionize the system through the energy transfer from the CM to
the internal motion according to Eq.(14).
Perpendicular to the magnetic field the muonic motion
is finite because of the confining property of the magnetic
field. This is a principal difference of the self-ionization in
the presence of a magnetic field compared to the classical ionization by an
external electric field.

In the present calculations (the initial population is
chosen to be $P_{15s}(t=0)=1$, i.e. the $n=15, l=m=0$ state)
we used a more detailed grid over $r\in [0,r_m=1500a_{0}]$
(400 grid points) as
compared to the calculations for the low-lying states $n=2,3$.
The boundary $r_m$ was chosen approximately 10 times exceeding the initial
value of the mean radius $\la r\ra$ of the $\mu He^+$ ion in the $n=15, l=m=0$
state. Following the Refs.\cite{15,29} we have used an absorbing
boundary condition at the point $r=r_m$. It permits to prevent the
artificial reflection of the muon flux from the grid boundary as well as
allows an estimation of the ionization rate by analyzing the decay of the muon
norm $N_{15s}(t)=\int\mid\psi(\ve{r},t)\mid^2 d\ve{r}$ with
time\cite{29}.

The calculated time-evolution of the muon densities in $z$-direction
of $\ve{B}$ and the perpendicular direction are
presented in Figs.4 for the strong
field $B=4\cdot 10^{3} Tesla$. They demonstrate
considerable spreading of the muon density in $z$-direction for $t\geq
2\cdot 10^{-11} s$. The evaluated norm $N_{15s}(t)$ (Fig.5) gives the
following estimate $t_I\simeq 10^{-10} s$ for the order of the
self-ionization time. This shows that the self-ionization is at least
one order of magnitude faster than the process of the muonic molecule resonant
formation ($\sim 10^{-8}-10^{-9} s$[1-4]) and potentially may be useful for
increasing the muon stripping from the highly-excited states ($n\sim15$) of
the $(\mu He^{+})_n$ ions. However, it is
at the time being not clear
how to make this process efficient
with respect to the increase of the reactivation rate
R. Actually, the $(\mu He^{+})_n$ ions are formed in reaction (11) mainly
in low-lying excited states (only a few percents of the muonic ions are in
$n>5$[1-4]) for which the self-ionization is much slower.
Our estimates, made for the same field 
strength $B=4\cdot 10^{3} Tesla$ 
show a fast increase of the self-ionization time with decreasing $n$.
Particularly, for $n=5$ the self-ionization time already exceeds the critical
value $\sim 10^{-8} s$ defined by the resonant muonic molecule formation.
In principle it may be compensated by increasing the strength
of the applied field to $B>4\cdot 10^{3} Tesla$ , which is, however,
orders of magnitude beyond the
currently available fields at high magnetic field facilities\cite{30}.
The so far kown mechanisms\cite{26} also do not permit an efficient
excitation of the fast muonic ions.

\section{Conclusions}

In this work we analyzed the transitions stimulated by external magnetic
fields from excited states of the fast $\mu He^{+}$ ions.
The results have been obtained within an approach developed for treating
non-adiabatic two-body charged system in external magnetic
fields on a mixed quantum-semiclassical level.

It was shown that the present-day laboratory fields of the order of a few
$Tesla$ may stimulate strong $2s-2p$ mixing as well as $l$ mixing of the
sub-levels for $n > 2$ in the $\mu He^{+}$ ions formating in $\mu CF$. The
effect of the magnetic field stimulating the $2s-2p$ transitions can be
analyzed experimentally by measuring the dependence of the intensities of
the $K_{\alpha}$-lines from the $\mu He^{+}$ on the field strength. It is
also interesting to analyze experimentally the influence of this effect on
another important $\mu CF$ parameter, the reactivation rate $R$, due to
the possible dependence of the value $R$ on the $2s-2p$ mixing in $\mu
He^+$\cite{24,25}.

The possibility of the self-ionization
process for fast $\mu He^{+}$ ions in
highly excited states has been demonstrated for strong magnetic fields.

V.S.M. thanks Dr. J.S.~Cohen for useful comments, Professor
W.H.~Breunlich and the staff of IMEP of the Austrian Academy of Sciences
for warm hospitality and the use of the computer resources.
This work was partially supported by the Belgian program on interuniversity
attraction poles (V.S.M.), initiated by the Belgian-state Federal Services
for Scientific, Technical and Cultural Affairs, and also by the
National Science Foundation through a grant (P.S.) for the Institute for
Theoretical Atomic and Molecular Physics at Harvard University and
Smithsonian Astrophysical Observatory.

\vspace*{2.0cm}

{\bf FIGURE CAPTIONS}\\
\\
Fig. 1: Magnetic field stimulating the transitions between the
$lm$ sub-levels of excited states $n=3$ and $2$ in the fast
$(\mu He^{+})_{nlm}$ ion. The calculation of the populations
$P_{nl}(t)$ was performed for the fixed projection $v_{\perp}=6
v_0$ of the $\mu He^{+}$ initial CM velocity on the plane
perpendicular to the magnetic field $\ve{B}$ for $B = 4 Tesla$.
The initial populations have been chosen as $P_{3s}(t=0)=1$ and
$P_{2s}(t=0)=1$.
\\

Fig. 2: Dependence of the time-evolution of the population $P_{2s}(t)$
on the magnetic field strength $B$ (for fixed $v_{\perp}=6\cdot v_0$)
and the $v_{\perp}$ component of the initial CM velocity $v$ of the
$\mu He^{+}$ ion (for fixed $B=2 Tesla$). The mean value
$\bar v_{\perp} =
\sqrt{2/3} \cdot 6 v_0\simeq 5\cdot v_0$
corresponds
to the initial kinetic energy $3.5 MeV$ of the $\mu He^{+}$ ion
emitting in $dt$ fusion reactions.
\\

Fig. 3: Dependence of the time-evolution of the population $P_{2s}(t)$
with and
without the relativistic splitting $\Delta E_{2S-2P}$. Broken curves have
been calculated without the VP term (15), the solid lines correspond to 
calculations including the VP term in the effective Hamiltonian (7).  \\

Fig. 4: Time-evolution of the muon densities
$[\psi(\rho=0,z,t)]^2$ and $[\psi(\rho,z=0,t)]^2$ starting with
the initially populated state $nlm =15,0,0$ for
$B=4\cdot 10^3 Tesla$. The distances are given in muonic atomic units
$a_0=2.56\cdot 10^{-11} cm$.\\

Fig. 5: Norm $N_{nlm}(t)$ decay of
the initially populated state $nlm =15,0,0$ for
$B=4\cdot 10^3 Tesla$. \\
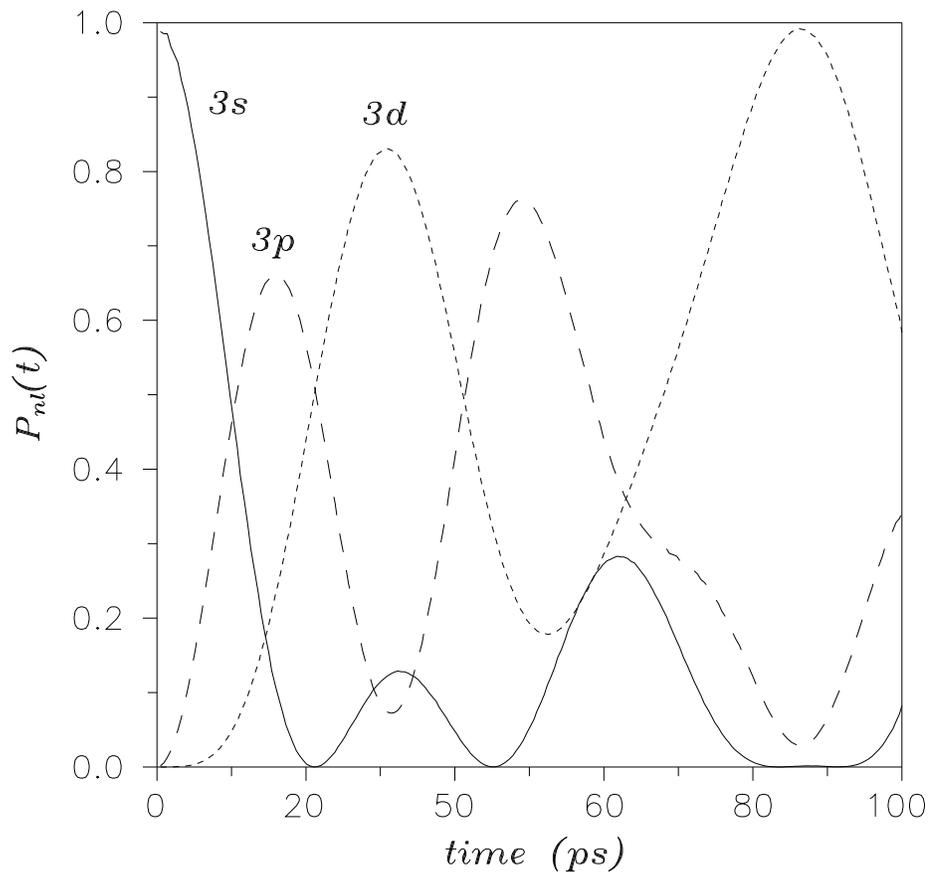
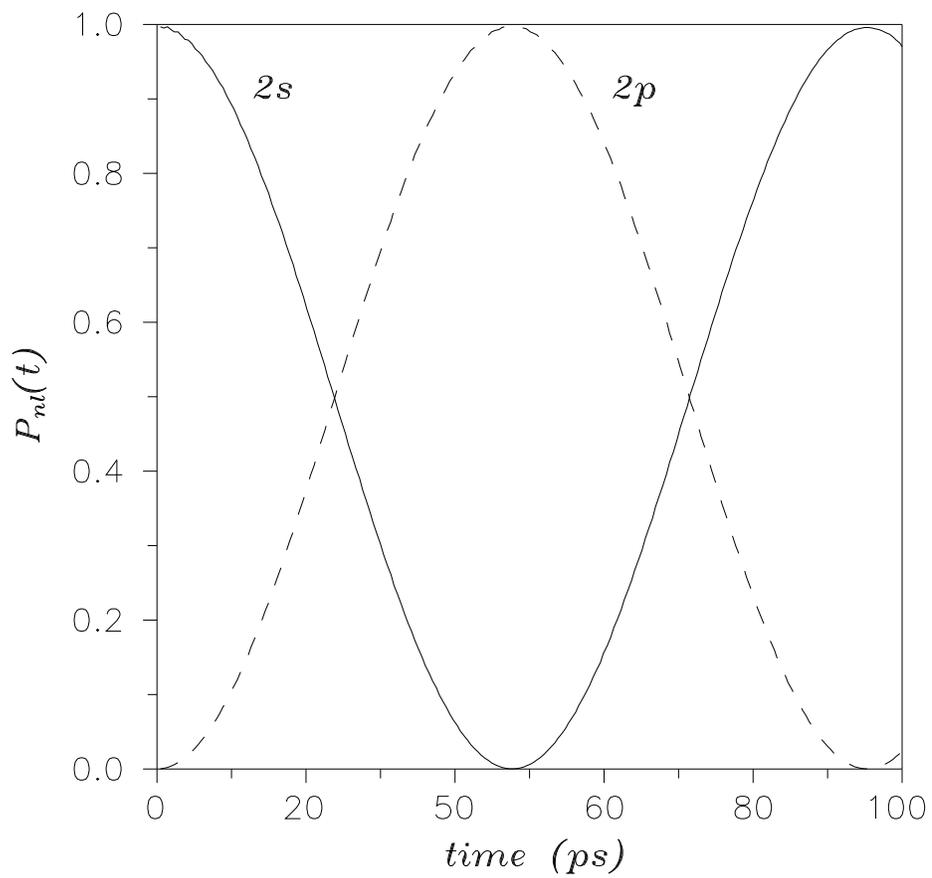
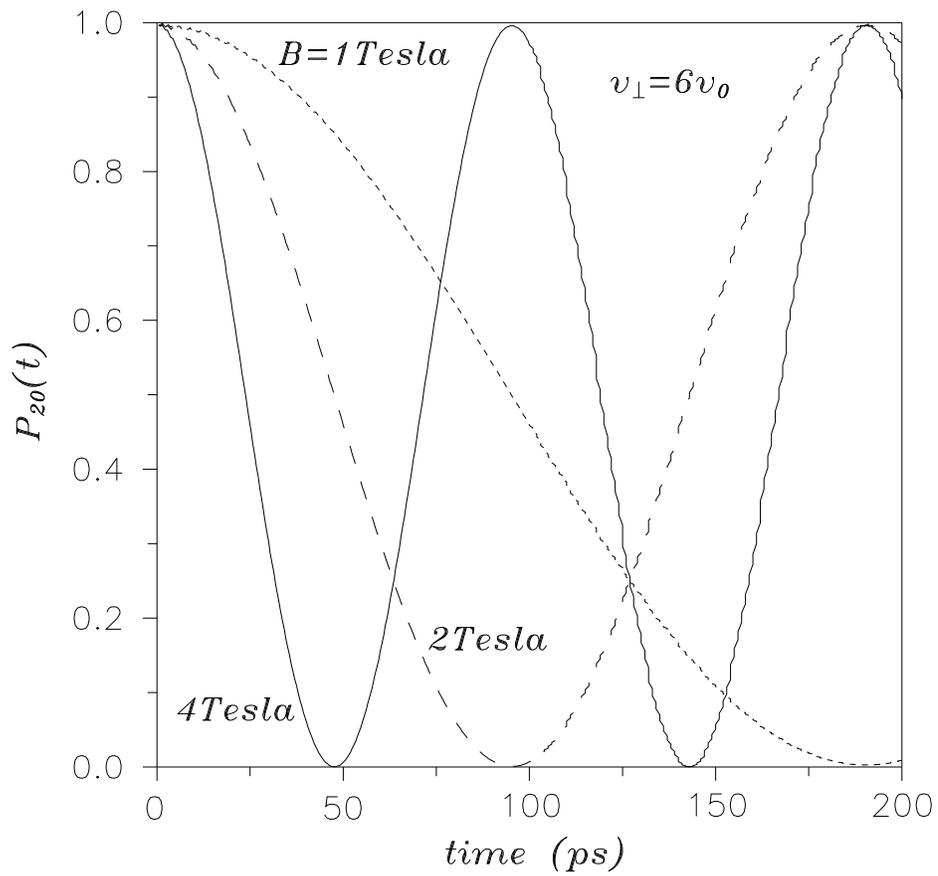
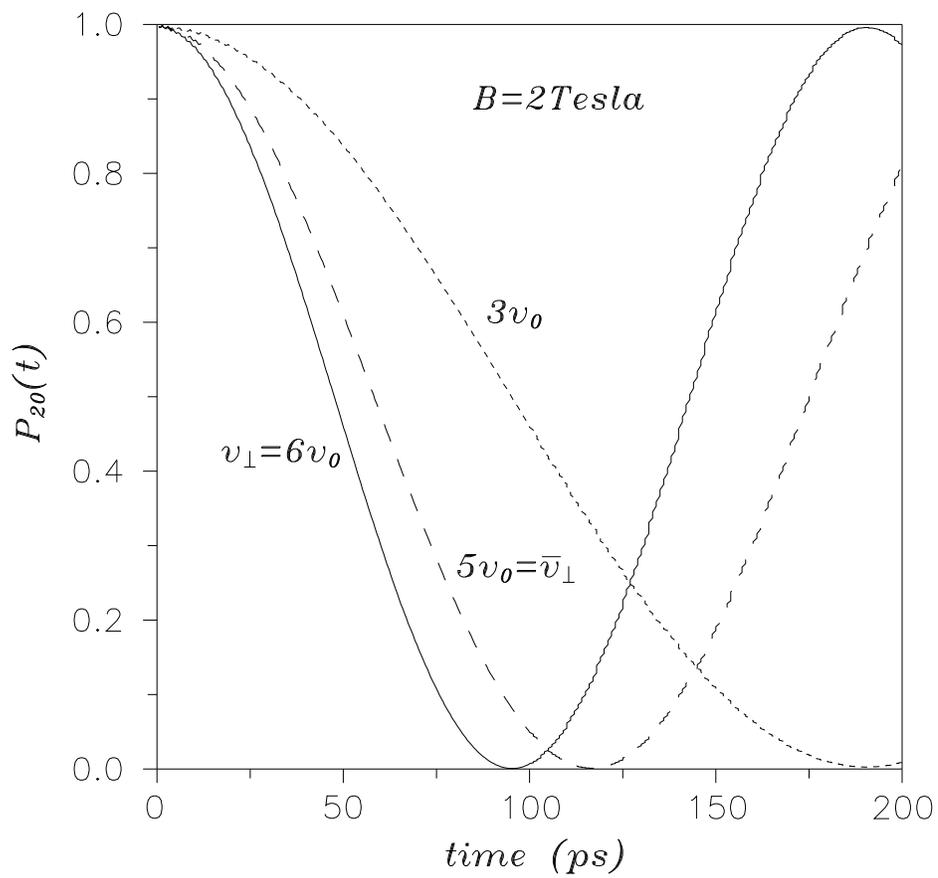
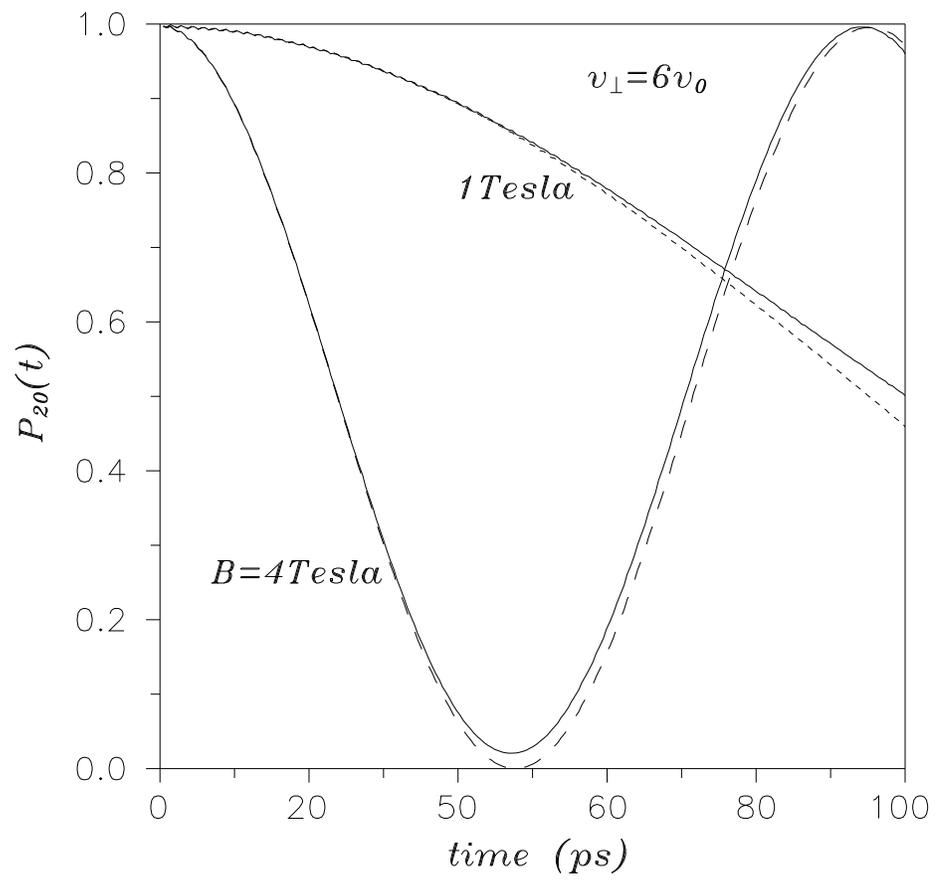
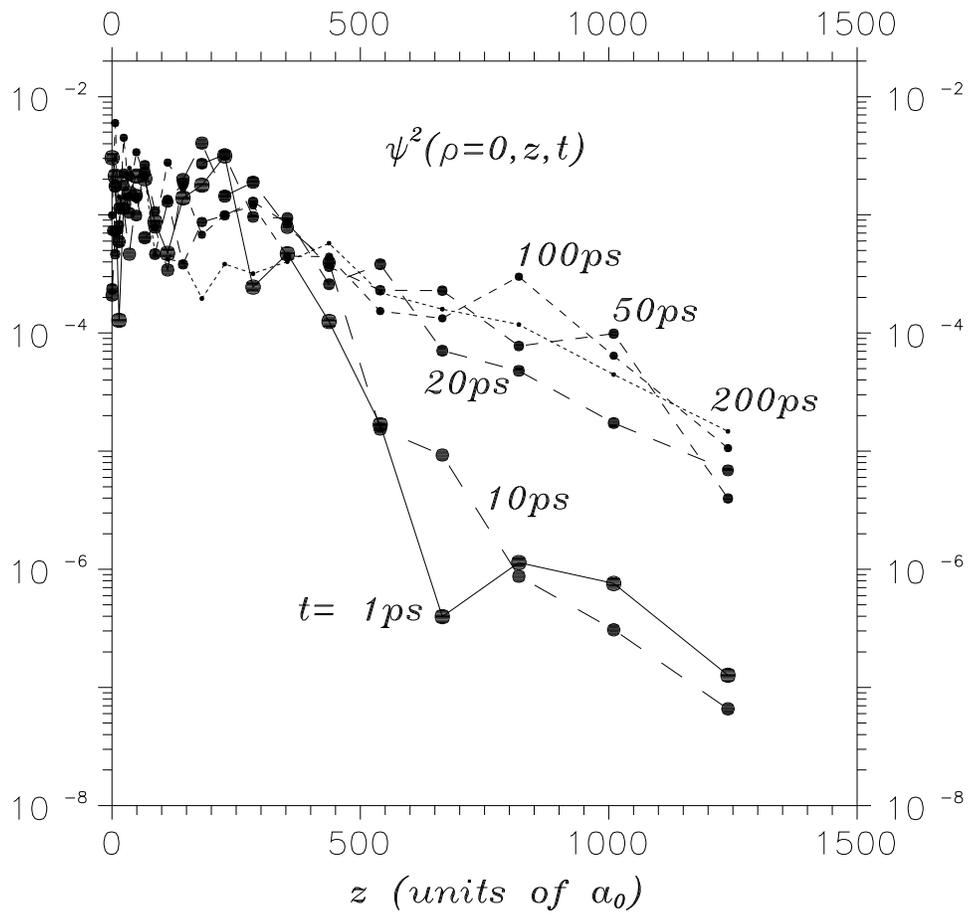
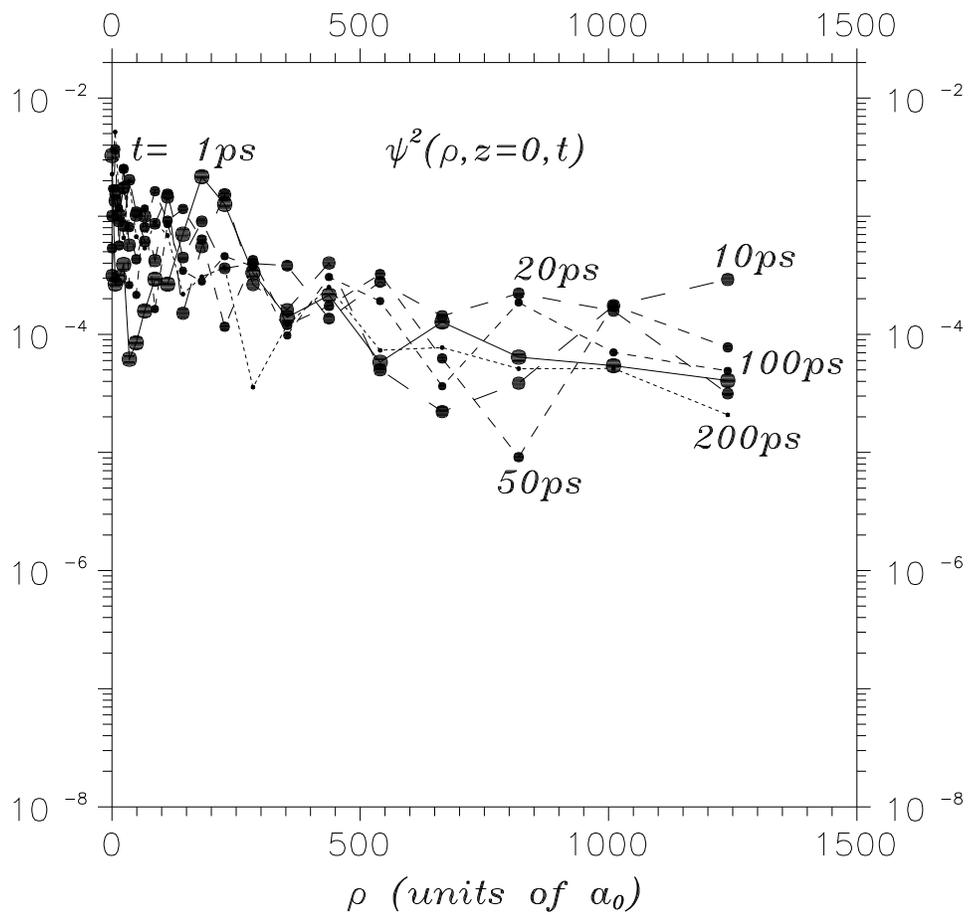
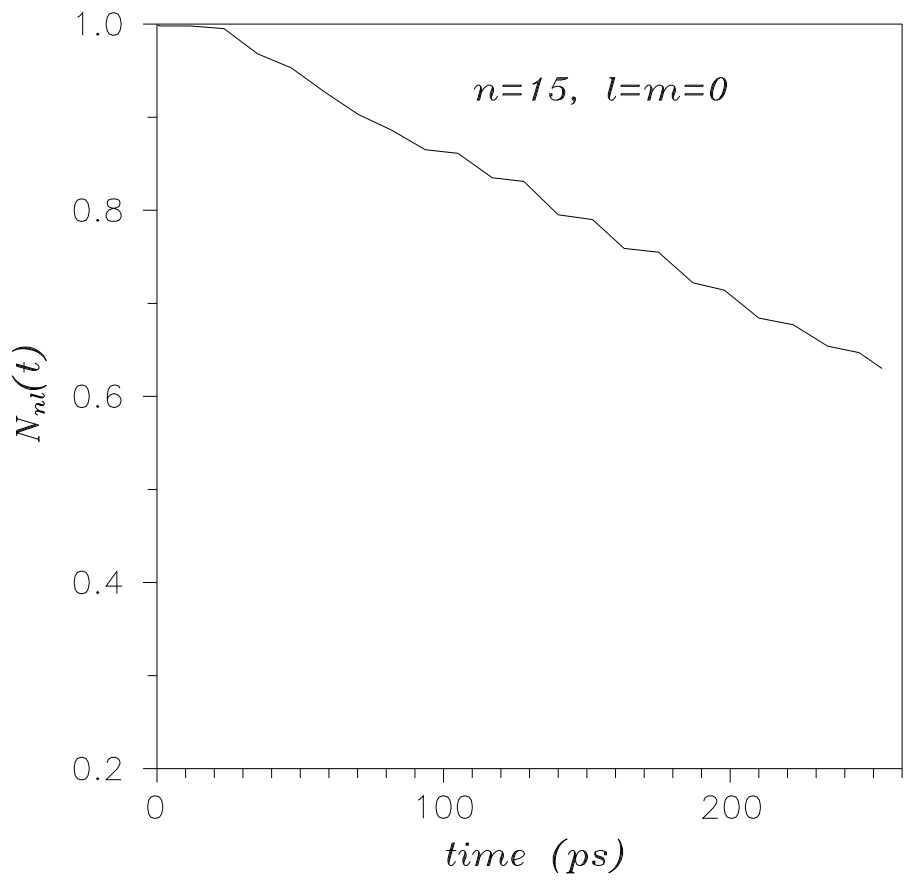
\end{document}